\documentclass[a4paper, 12pt, fleqn]{article}
\begin{document}
\begin{center}
\textbf{\large ENVIRONMENT-DEPENDENT FUNDAMENTAL PHYSICAL CONSTANTS}
\end{center}
\begin{center}
\textbf{\large $-$In the Theory of General Inconstancy}

\vspace{5mm}
Hidezumi Terazawa
\footnote{E-mail address: \textit{terazawa@mrj.biglobe.ne.jp} }
\end{center}

\vspace{0mm}
\begin{center}
\textit{Center of Asia and Oceania for Science(CAOS)}

\textit{3-11-26 Maesawa, Higashi-kurume, Tokyo 203-0032, Japan}

\textit{and}

\textit{Midlands Academy of Business \& Technology(MABT),}

\textit{Mansion House, 41 Guildhall Lane, Leicester LE1 5FR, United Kingdom}

\end{center}
\begin{center}
\textbf{Abstract}
\end{center}

A theory of special inconstancy, in which some fundamental physical constants such as 
the fine-structure and gravitational constants may vary, is proposed in pregeometry. 
In the special theory of inconstancy, the $\alpha-G$ relation of $\alpha=3\pi/[16\ln(4\pi/5GM_W^2)]$ between 
the varying fine-structure and gravitational constants (where $M_W$ is the charged 
weak boson mass) is derived from the hypothesis that both of these constants are 
related to the same fundamental length scale in nature. Furthermore, it leads to the 
prediction of $\dot{\alpha}/\alpha=(-0.8\pm2.5)\times10^{-14}yr^{-1}$ from the most precise limit of 
$\dot{G}/G=(-0.6\pm2.0)\times10^{-12}yr^{-1}$ by Thorsett, which is not only consistent with 
the recent observation of $\dot{\alpha}/\alpha=(0.5\pm0.5)\times10^{-14}yr^{-1}$ by Webb \textit{et al.} 
but also feasible for future experimental tests. In special inconstancy, the past and present 
of the Universe are explained and the future of it is predicted, which is quite different 
from that in the Einstein theory of gravitation. Also, a theory of general inconstancy, in which any 
fundamental physical constants may vary, is proposed in \lq\lq more general relativity\rq\rq, by assuming that 
the space-time is \lq\lq environment-dependent\rq\rq. In the general theory of inconstancy, 
the $G-\Lambda$ relation between the varying gravitational and cosmological constants is derived from the hypothesis 
that the space-time metric is a function of $\tau$, the \lq\lq environment coordinate\rq\rq, 
in addition to $x^{\mu}$, the ordinary space-time coordinates. Furthermore, it leads to the prediction of 
the varying cosmological constant, which is consistent with the present observations. In addition, the latest 
observation of spatial variation in the fine-structure constant from VLT/UVES of $(1.1\pm 0.2)\times 10^{-6}
GLyr^{-1}$ by King \textit{et al.} is suggested to be taken as a clear evidence for environment-dependent 
fundamental physical constants.

\vspace{10mm} 
\textbf{\large I. Introduction}

\vspace{5mm}
Is a physical constant really constant? In 1937, Dirac[1] discussed possible time variation in the 
fundamental constants of nature. He made not only the large number hypothesis(LNH) but also, as a consequences of the 
LNH, the astonishing prediction that the gravitational constant $G$ varies as a function of time. Since then, 
Jordan[2] and many others[3,4] have tried to construct new theories of gravitation or general relativity in order 
to accomodate such a time-varying $G$. Although the LNH has been inspiring many theoretical developements and has 
recently led myself[5] to many new large number relations, the prediction of the varying $G$ has not yet received any 
experimental evidence. Recently, Thorsett[6] has shown that measurements of the masses of young and old neutron stars 
in pulsar binaries lead to the most precise limti of \[\dot{G}/G=(-0.6\pm2.0)\times10^{-12}yr^{-1}\] at the 68\% 
confidence level.

More recently, on the other hand, Webb \textit{et al.}[7] have investigated possible time variation in 
the fine structure constant $\alpha$ by using quasar spectra over a wide rage of epoches, spanning redshifts $0.2<z<3.7$, 
in the history of our Universe, and derived the remarkable result of \[\dot{\alpha}/\alpha=(6.40\pm1.35)\times10^{-16}
yr^{-1}\] for $0.2<z<3.7$, which is consistent with a time-varying $\alpha$. Note, however, that in 1976 Shylakhter[8] 
obtained the very restrictive limit of $|\Delta\alpha/\alpha|<10^{-7}$ or, more precisely, \[\Delta\alpha/\alpha\Delta 
t=(-0.2\pm0.8)\times10^{-17}yr^{-1}\] for $z\sim 0.16$ (but over a narrower and latest range of epochs between now and 
about 1.8 billion years ago) from the \lq\lq Oklo natural reactor\rq\rq. Very lately, Srianand \textit{et 
al.}[9] have made a detailed many-multiplet analysis performed on a new sample of Mg II systems observed in high quality 
quasar spectra obtained using the Very Large Telescope and found a null result of $\Delta\alpha/\alpha=(-0.06\pm0.06)
\times10^{-5}$ for the fractional change in $\alpha$ or $3\sigma$ constraint of \[-2.5\times10^{-16}yr^{-1}\leq 
(\Delta\alpha/\alpha\Delta t)\leq+1.3\times10^{-16}yr^{-1}\] for $0.4\leq z\leq 2.3$, which seems to be inconsistent 
with the result of Webb \textit{et al.}[7]. However, a careful comparison of these different results[7-8] indicates that 
they are all consistent with a time-varying $\alpha$ as \[\dot{\alpha}/\alpha=(0.5\pm0.5)\times10^{-14}yr^{-1}\] for 
$2.2<z<3.7$. More lately, Kanekar \textit{et al.} have constrained the fundamental constant evolution with HI and OH lines 
as $\Delta\alpha/\alpha=(-1.7\pm1.4)\times10^{-6}$ over a lookback time of 6.7Gyrs or, equivalently, \[\dot{\alpha}/\alpha
=(-2.5\pm2.1)\times10^{-16}yr^{-1}\] for $0\leq z\leq 0.765$[8].

In this paper, I am going to propose a theory of special inconstancy, in which some fundamental physical 
constants such as the fine-structure and gravitational constants may vary. In the special theory of inconstancy, the 
$\alpha-G$ relation of \[\alpha=3\pi/[16\ln(4\pi/5GM_W^2)]\](where $M_W$ is the charged weak boson mass) is derived 
from the hypothesis that both of $\alpha$ and $G$ are related to the same fundamental length scale in nature. 
Furthermore, from the limit on $\dot{G}$ by Thorsett, it leads to the prediction of \[\dot{\alpha}/\alpha=(-0.8\pm2.5)
\times10^{-14}yr^{-1}\]which is not only consistent with the above result on $\dot{\alpha}$ but also feasible for 
future experimental tests. In special inconstancy, the past and present of the Universe are explained and the future of 
it is predicted, which is quite different from that in the Einstein theory of gravitation. Also, a general theory of 
inconstancy, in which any fundamental physical constants may vary, is proposed in \lq\lq more general relativity\rq\rq, 
by assuming that the space-time is \lq\lq environment-dependent\rq\rq. In the general theory of inconstancy, the 
\lq\lq $G-\Lambda$ relation\rq\rq\ between the varying gravitational and cosmological constants is derived from the hypothesis 
that the space-time metric is a function of $\tau$, the \lq\lq environment coordinate\rq\rq, in addition to $x^{\mu}$, 
the ordinary space-time coodinates. Furthermore, it leads to the prediction of the varying cosmological constant, 
which is consistent with the present experimental observations. 

I will organize this paper as follows: in Section II, I will briefly review pregeometry in which a theory of special 
inconstancy is constracted. In Sections III and IV, I will present the theories of special and general inconstancy 
and their predictions, respectively. Finally in Section V, I will discuss future prospects. 

\vspace{5mm}
\textbf{\large II. Pregeometry}

\vspace{5mm}
Pregeometry is a theory in which Einstein geometrical theory of gravity in general relativity can be derived from 
a more fundamental principle as an effective and approximate theory at low energies (or at long distances). In 1967, 
Sakharov[10] suggested possible approximate derivation of the Einstein-Hilbert action from quantum fluctuations of matter. 
A decade later, we[11] demonstrated that not only Einstein theory of gravity in general relativity but also 
the standard model of strong and electroweak interactions in quantum chromodynamics and in the unified gauge theory 
can be derived as an effective and approximate theory at low energies from the more fundamental unified 
composite model of all fundamental particles and forces[12].

Let us explain what pregeometry means more explicitly in a simple model of \[S_0=\int d^4x\surd\overline{-g}L_0
(g_{\mu\nu}(x), A_{\mu}(x), \phi_i(x))\] where $g_{\mu\nu}$ is the space-time metric, $g=det(g_{\mu\nu})$, $A_{\mu}$ is an 
Abelian gauge field, and $\phi_i$ ($i=1\sim n$) are $n$ complex scalar fields of matter with the charge $e$. The fundamental 
Lagrangian $L_0$ consists of the gauge-invariant kinetic terms of the matter fields only as \[L_0=g^{\mu\nu}
[\partial_{\mu}+iA_{\mu})\phi_i^{\dagger}](\partial_{\nu}-iA_{\nu})\phi_i-F^{-1}\] (where $F$ is an arbitrary constant) but 
does not contain either the kinetic term of the space-time metric or that of the gauge field so that both of $g_{\mu\nu}$ 
and $A_{\mu}$ are auxiliary fields. The effective action for the space-time metric and gauge field can be defined by 
the path-integral over the matter fields as \[exp(iS_{eff})=\int \prod_i[d\phi_i^{\dagger}][d\phi_i]exp(iS_0)\] and it can 
be expressed formally as \[S_{eff}=-iTr ln[(\partial_{\nu}-iA_{\nu})\surd\overline{-g}g^{\mu\nu}(\partial_{\mu}+iA_{\mu})]
-\int d^4x\surd\overline{-g}F^{-1}\]after the path-integration over $\phi_i$. For small scalar curvature $R$ and Ricci 
curvature tensor $R_{\mu\nu}$, the effective action can be calculated to be \[S_{eff}=\int d^4x\surd\overline{-g}
[2\lambda+(1/16\pi G)R+c(R^2+dR^{\mu\nu}R_{\mu\nu})+(1/4e^2)F^{\mu\nu}F_{\mu\nu}+...]\] with 
\[2\lambda=[n\Lambda^4/8(4\pi)^2]-F^{-1},\]\[(1/16\pi G)=n\Lambda^2/24(4\pi)^2,\]\[c=nln \Lambda^2/240(4\pi)^2,\] 
\[d=2,\] and \[(1/4e^2)=nln \Lambda^2/3(4\pi)^2,\] where $\lambda$ and $\Lambda$ are the cosmological constant and 
the momentum cut-off of the Pauli-Villars type, respectively. Note that the arbitrary constant $F^{-1}$ plays a role of 
counter term so that the cosmological constant may become as small as it is observed. Note also that the momentum cut-off 
$\Lambda$ must be of order of the Planck mass $G^{-1/2}$($\sim 10^{19}GeV$). Furthermore, not only the $R^2$ and 
$R^{\mu\nu}R_{\mu\nu}$ terms but also the remaining terms in the expansion of $S_{eff}$ are practically negligible. 
This complete a simple demonstration that not only the Einstein-Hilbert action of gravity but also the Maxwell action of 
electromagnetism in general relativity can be derived as an effective and approximate theory at low energies from 
the simple model in pregeometry, provided that there exists a natural momentum cut-off at around the Planck mass 
in nature[13].

One of the most remarkable consequences of pregeometry is the $\alpha-G$ relation, a simple relation between 
the fine-structure and gravitational constant, which can be easily derived from the results for $\alpha$ and for 
$G$ by eliminating the momentum cut-off $\Lambda$. In our unified quark-lepton model of all fundamental forces[14,15], 
the $\alpha-G$ relation is given by[16] \[\alpha=3\pi/\sum_i Q_i^2 ln(12\pi/nGm_i^2),\] where $Q_i$ and $m_i$ are 
the charge and mass of quarks and leptons, respectively. For three generations of quarks and leptons and their 
mirror- or super-partners, the $\alpha-G$ relation simply becomes \[\alpha\cong 3\pi/16ln(4\pi/5GM_W^2)\] 
where $M_W$ is the charged weak boson mass. Note that this $\alpha-G$ relation is very well satisfied by the experimental 
data of $\alpha\cong 1/137$, $G^{-1/2}\cong 1.22\times10^{19}GeV$, and $M_W\cong 80.4GeV$[17].

\vspace{5mm}
\textbf{\large III. Special Inconstancy}

\vspace{5mm}
Special inconstancy is a principle in which some fundamental physical constants such as the fine-structure and gravitational 
constants may vary. Let us first make it clear that in this paper we use the natural unit system of $h/2\pi=c=1$ 
(where $h$ is the Planck constant and $c$ is the speed of light in vacuum). Note, however, that it does not mean that 
in discussing the relevant possibility of the varying fine-structure and gravitational constants[18], 
we exclude another intriguing possibility of the varying light velocity recently discussed by some authors[19] 
since varying either $h$ or $c$ is inevitably related to varying the fine-structure constant $\alpha$
($\equiv e^2/2hc$)(if the unit charge $e$ stays constant). It simply means that we must set up a certain reference 
frame on which we can discuss whether physical quantities such as the fine-structure, gravitational, and cosmological[20] 
constants be really constant. Our basic hypothesis is that both of the fine-structure and gravitational constants are 
related to the more fundamental length scale of nature as in the unified (pregauge[21] and) pregeometric[10-12] theory 
(or \lq\lq pregaugeometry\rq\rq\ in short) of all fundamental forces[14,15] reviewed in the last Section.

To be more explicit, in the simple model of pregaugeometry discussed in the last Section, assert that 
\[<(\partial_{\mu}+iA_{\mu})\surd\overline{-g}g^{\mu\nu}(\partial_{\nu}-iA_{\nu})\phi_i>_{\Lambda}=0,\]
\[g_{\mu\nu}=F<[(\partial_{\mu}+iA_{\mu})\phi_i^{\dagger}](\partial_{\nu}-iA_{\nu})\phi_i>_{\Lambda},\] and 
\[A_{\mu}=(i/2)<[\phi_i^{\dagger}\partial_{\mu}\phi_i-(\partial_{\mu}\phi_i^{\dagger})\phi_i]/(\phi_j^{\dagger}
\phi_j)>_{\Lambda},\] where $<>_{\Lambda}$ denotes the expectation value in the space-time with the fundamental 
length scale parameter of $\Lambda^{-1}$. The first equation is the usual field equation for $\phi_i$ 
while the last two can be taken either as the \lq\lq equations of motion\rq\rq\ for $g_{\mu\nu}$ and $A_{\mu}$, 
which can be derived from the fundamental action $S_0$, or as the \lq\lq fundamental field equations\rq\rq, 
which can reproduce the effective Einstein-Hilbert-Maxwell action $S_{eff}$ at low energies($\ll \Lambda$) 
or at long distances($\gg \Lambda^{-1}$).

The most important consequence of special inconstancy in pregaugeometry is the $\dot{\alpha}-\dot{G}$ relaton for the 
varying fine-structure and gravitational constant of \[\dot{\alpha}/\alpha^2=(16/3\pi)[(\dot{G}/G)+2(\dot{M_W}/M_W)],\] 
which can be derived from differentiating both hand sides of the $\alpha-G$ relation with respect to any parameter 
for varying fundamental physical constants. This immediately leads to the remarkable predictions of 
\[\dot{\alpha}/\alpha=(-0.8\pm2.5)\times10^{-14}yr^{-1}\] for constant $M_W$ and 
\[\dot{M_W}/M_W=(0.5\pm1.2)\times10^{-12}yr^{-1}\] from the limit of $\dot{G}/G=(-0.6\pm2.0)\times10^{-12}yr^{-1}$ by 
Thorsett[6] and from the experimental data of $\dot{\alpha}/\alpha=(0.5\pm0.5)\times10^{-14}yr^{-1}$ by Webb 
\textit{et al.}[7]. The first prediction is not only consistent with the experimental data by Webb \textit{et al.}[7] 
but also feasible for future experimental tests. The second prediction, however, seems too small to be feasible 
for experimental tests in the near future although such prediction for the possible varying particle masses seems 
extremely interesting at least theoretically. Note that the varying $M_W$ is perfectly possible 
through the varying electroweak gauge coupling constant $g$ (which is related to the fine-structure constant 
in the standard unified electroweak gauge theory of Glashow-Salam-Weinberg[22]) and/or the varying vacuum 
expectation value of the Higgs scalar $v$ (which is related to the momentum cut-off $\Lambda$ in the unified 
composite model of the Nambu-Jona-Lasinio type for all fundamental forces[23]) since $M_W=gv/2$. We must mention, 
however, that neither one of these predictions may assert for the fine-structure constant $\alpha$ or 
the weak boson mass $M_W$ to vary since both of these predicted values are consistent with zero within their errors.

Let us now first add that in some pregaugeometric model[24], the $\alpha-G$ relation is not of the type of $\alpha\sim 
1/ln(1/GM^2)$ but of the type of $\alpha\sim GM^2$(where M is a parameter of mass dimension) so that the 
$\dot{\alpha}$-$\dot{G}$ relation becomes \[\dot{\alpha}/\alpha=(\dot{G}/G)+2(\dot{M}/M).\] This type of relation 
predicts \[\dot{\alpha}/\alpha=(-0.6\pm2.0)\times 10^{-12}yr^{-1}\] for constant $M$ and \[\dot{M}/M=(0.3\pm1.0)
\times10^{-12}yr^{-1}\] from the limit by Thorsett[6] and from the experimental data by Webb \textit{et al.}[7]. 
We suspect, however, that the predicted value for $\dot{M}/M$ seems too small to be feasible for experimental tests 
in the near future although the one for $\dot{\alpha}/\alpha$ is consistent with the experimental data by Webb 
\textit{et al.}[7]. We must also mention that both of these predicted values are consistent with zero within 
their errors so that neither the fine-structure constant $\alpha$ nor the mass parameter $M$ may be predicted to vary.

Next, remember that in the principle of special inconstancy we do not assert that physical constants may vary 
as a function of time but do that they may vary in general, depending on any parameters including the cosmological time, 
temperature, \textit{etc.}. What is the origin of varying the physical constants? The answer to this question may be 
related to the answer to another fundamental question: What is the origin of the fundamental length scale 
$\Lambda^{-1}$ in nature? It can be spontaneous breakdown of scale-invariance in the Universe, which has been 
proposed by myself[5] for the last quater century. It can be the natural, dynamical, automatic, \textit{a priori}, 
but somewhat \lq\lq wishful-thinking\rq\rq\ cut-off at around the Planck length $G^{1/2}$ where gravity would become 
as strong as electromagnetism, which was suggested by Landau[13] in 1955. It can also be due to the Kaluza-Klein 
extra dimension[25], which is supposed to be compactified at an extremely small length scale of the order of $G^{-1/2}$ 
or at a relatively large length scale of the order of $1/TeV$ recently emphasized by Arkani-Hamed \textit{et al.}[26]. 
It seems, however, the most natural and likely that the origin of the fundamental length comes from the substructure 
of fundamental particles including quarks, leptons, gauge bosons, Higgs scalars, \textit{etc.}[27,28]. In the unified 
composite model of all fundamental particles and forces[28], the fundamental energy scale $\Lambda$ in pregaugeometry 
can be related to some even more fundamental parameters such as the masses of subquarks, the more fundamental 
constituents of quarks and leptons, and the energy scale in quantum subchromodynamics, the more fundamental dynamics 
confining subquarks into a quark or a lepton. In either way, the fundamental length scale $\Lambda^{-1}$ can be 
identified with the size of quarks and leptons, the fundamental particles.

In pregeometric special inconstancy, let us now briefly explain the past and present of the Universe and try to predict 
the future of it, which may differ from that in the conventional Einstein theory of gravitation in general relativity[29]. 
The history of our Universe goes as follows: Long, long time ago there was no physical space-time, in which the 
space-time metric was finite and non-vanishing so that the distance was well defined, but the only matter 
\lq\lq existed\rq\rq\ in the mathematical space-time. Suddenly, there appeared the big bang of our Universe 
as a phase transition of the space-time from the pregeometric phase to the geometric one due to quantum fluctuations 
of matter, as suggested by us[30] in the early nineteen eighties, and our Universe had happened to be either flat 
or open. Then, not only all fundamental particles but also all fundamental forces between them were created and 
they started obeying the effective theory of all fundamental particles and forces including the Einstein theory of 
gravity with the non-vanishing and varying cosmological constant. In the earliest era during which the matter 
density had been extremely small, our Universe had been expanding almost exponentially. It had been the \lq\lq almost 
inflationary Universe\rq\rq. In the next era of the radiation dominated Universe, our Universe was expanding less fast. 
Furthermore, in the last era of the matter dominated Universe, our Universe has still been expanding even faster. 
This history of our Universe is well simulated by a simple model of $(\Omega_m, \Omega_{\lambda}, -q)=(0,1,1)$, 
$(1/4,3/4,1/3)$, or $(1/4,3/4,1/2)$ for the early inflationary era, for the radiation dominated era, or for the 
matter dominated era, respectively, where $\Omega_m$, $\Omega_{\lambda}$, and $q$ are the \lq\lq 
pressureless-matter-density\rq\rq, \lq\lq scaled cosmological constant\rq\rq, and deceleration parameter of the Universe, respectively. 
Note that there must be another \lq\lq phase transition\rq\rq\ in which $\Omega_{\lambda}$ changed from 1 to 2/3 
in between the early inflationary era and the radiation dominated era. Concerning the cosmological constant, 
I have been most impressed by the recent observation of the \lq\lq farthest supernova ever seen\rq\rq\ by Hubble 
Space Telescope[31]. \lq\lq This supernova shows us the universe is behaving like a driver who slows down 
approaching a red stoplight and then hits the accelerator when the light turns green.\rq\rq\ Note that this behavior 
of the Universe is what our model simulates. Note also that our model of the Universe is consistent with the recent 
measurement of the cosmological mass density from clustering in the Two-Degree-Field Galaxy Redshift Survey[32] 
which strongly favors a low density Universe with $\Omega_m\cong 0.3$. Very recently, the CBI Collaboration[33,17] has 
found $\Omega_{\lambda}=0.64+0.11/-0.14$ and $\Omega_m+\Omega_{\lambda}=1.006\pm 0.006$. More recently, the Wilkinson 
Microwave Anisotropy Probe (WMAP) team[34,17] has found 1) the first generation of stars to shine in the Universe first 
ignited only 200 million years after the big bang, 2) the age of the Universe is $13.69\pm 0.13$ billion years old, and 
3) $\Omega_m=0.26\pm 0.02$ and $\Omega_{\lambda}=0.74\pm 0.03$.

The future of the Universe in our special inconstant picture can be quite different from that in the 
Einstein-Friedmann picture: 1) Since the cosmological constant may vary in special inconstancy, the space-time 
of our Universe which is almost flat and expanding faster and faster may not continue to be flat and accelerating 
forever. Our Universe may even encounter a \lq\lq topological phae transition\r\rq, which was first discussed 
by Wheeler[35] in 1959, from the open Universe to the closed one. 2) If the gravitational constant increases, the 
expansion of the space-time may not continue forever. The Universe may well stop expanding, start contracting, 
and even be bouncing forever. If $G$ decreases, it will be more accelerated ever. 3) If the fine-structure constant 
(and/or other fundamental coupling constants such as the strong and weak coupling constants) varies, our Universe 
may encounter as \lq\lq obsolete phase transition\rq\rq\ from the matter-dominated Universe to the radiation-dominated 
one. In short, we can expect anything about the future of our Universe or, in other words, we can predict nothing 
definite on the destiny of our Universe.

\vspace{5mm}
\textbf{\large IV. General Inconstancy}

\vspace{5mm}
General inconstancy is a principle in which any fundamental physical constants may vary. In order to accomodate general 
inconstancy with general relativity, let us assume that the space-time is environment dependent. More explicitly, let us 
make the hypothesis that the space-time metric, $g_{\mu\nu}$, is a function of the \lq\lq environment coordinate\rq\rq, 
$\tau$, in addition to the ordinary space-time coodinates, $x^{\mu}$, where $\mu,\nu=0,1,2,3$, so that the infinitesimal 
distance, $ds$, is given by \[
ds^2=g_{\mu\nu}(x,\tau)dx^{\mu}dx^{\nu}.\] Note that the physical meaning of $\tau$ is arbitrary as it can be the 
temperature($T$), the cosmological time($t$), a parameter related to the Kaluza-Klein extra dimensions, or anything else. 
Note also that $\tau$ may not be a continuous valuable but be a discrete number and that there may exist more than one 
\lq\lq environment coordinates\rq\rq. I will consider such extensions of the environment-dependent space-time metric later.

In this \lq\lq more general relativity\rq\rq, let us 
consider a simple model for general inconstancy given by the generalized Einstein-Hilbert action of \[
S_I=\int d\tau \int d^4x\surd\overline{-g}[R(x,\tau)+...]=\int d^4x\surd\overline{-g}[2\lambda+(1/16\pi G)R(x)+...].\] 
In this equation we may find the equations of \[
2\lambda=\int d\tau \int d^4x\surd\overline{-g}[R(x,\tau)+...]/\int d^4x\surd\overline{-g}\] and \[
(1/16\pi G)=\int d\tau \int d^4x\surd\overline{-g}[R(x,\tau)+...]/\int d^4x\surd\overline{-g}R(x)\] so that we may obtain 
the $G-\Lambda$ relation of \[
32\pi G\lambda=\int d^4x\surd\overline{-g}R(x)/\int d^4x\surd\overline{-g}.\]
Futhermore, by differentiating both hand-sides of the relation with respect to $\tau$, we can obtain the relation between 
the varying gravitaional and cosmological constants:\[
\dot{G}/G=-\dot{\lambda}/\lambda.\]
Since in the homogeneous and isotropic Einstein-Friedmann model of the Universe, the scalar curvature is given by $R=
6[(1-q)H^2+(k/a^2)]$, where $a$ is the scale parameter in the Robertson-Walker metric and $k=+1,0,-1$ for the closed, 
flat, or open universe, the first relation explains the positivity of the cosmological constant, $\lambda$, for the 
present Universe with $-q>0$ and, probably, $k=0$. The second relation leads to the prediction of \[
\dot{\lambda}/\lambda=(0.6\pm2.0)\times 10^{-12}yr^{-1},\]
from the limit by Thorsett[6]. This prediction is not only consistent with the present experimental observation but also 
feasible for future experimental tests. 

I must mention, however, that I have not completed checking whether all the 
possible consequences of this simple theory of general inconstancy be consistent with all the presently available 
experimental observations. It is certainly the most important subject for future investigations. I have not found either 
whether it is necessary to extend it for discrete $\tau$ or for more than one \lq\lq environment coordinates\rq\rq. Just imagine 
that we stand in front of the entrance gate for the vast unkown world of general inconstancy!

\vspace{5mm}
\textbf{\large V. Future Prospects}

\vspace{5mm}
In conclusion, let us point out that not only continuous physical constants such as $\alpha$ and $G$ 
but also discrete physical numbers such as the number of the space-time dimensions $n$, the number of quark colors 
$N_c$, the number of quark-lepton generations $N_g$, \textit{etc.} may vary. In fact, an astonishing 
\lq\lq dimensional phase transition\rq\rq, which was discussed by myself[36] about two decades ago, may be possible 
in the history of our Universe. If $n$ is related to $N_c$ as in the \lq\lq space-color corespondence\rq\rq, 
which was proposed by myself about three decades ago[37], both of these fundamental physical natural numbers must 
vary simultaneously. Before concluding this talk, let me ask the following question: Are no constants of nature constant? 
After all, it may be that nothing is constant or permanent in the Universe as emphasized by the Greek and Indian 
philosophers about two and a half millennia ago!

In ending this paper, let us introduce two more recent reports on finding an evidence for environment-dependent fundamental 
physical constants in addition to many recent reports[38]. Very lately, Kanekar \textit{et al.} have obtained 
a very strong constraint of $\Delta\alpha/\alpha = (-1.7\pm 1.4)\times 10^{-6}$ over a lookback time of 6.7Gyrs 
with HI and OH lines[39]. Even more lately, King \textit{et al.} have found spatial variation of the fine-structure 
constant with the amplitude of $(1.1\pm 0.2)\times 10^{-6}GLyr^{-1}$ by using UVES on the VLT[40], which may be taken 
as a clear evidence for the environment-dependent fundamental physical constant. It may be a big-bang in physics!

\vspace{5mm}
\textbf{\large Acknowledgement}

\vspace{5mm}
The author would like to thank Professor Laszlo L. Jenkovszky for suggesting publication of 
this reduced and up-dated version of the original paper in Ukrainian Journal of Physics. In addition, 
he thanks Professor Yuichi Chikashige for very useful helps in converting the original manuscripts of this paper 
into those for publication in this Journal. This paper has been dedicated to Dr. Akira Tonomura, my late friend, 
who first observed the Aharonov-Bohm effect.
\end{document}